# Revisiting the Capacitance of a Hollow Metal Cylinder

Antonio Iadicicco $(a)$, Luigi Verolino $(b)$

$(a)$ Università popolare Nikola Tesla

Via Trinità 33, 83100 Avellino, Italy

$(b)$ DIETI, Università Federico II di Napoli

Via Claudio 21, 80125 Naples, Italy

Correspondence: antonio.iadicicco@unitesla.eu

We complete the analysis of the capacitance of a hollow metal cylinder by deriving series representations for different aspect ratios and a practical formula that approximates the capacitance of the tube for any aspect ratio with an error below 1%.

## 1. Introduction

We revisit a method [1] developed in 1995 for efficiently computing the capacitance of a hollow metal cylinder. Figure 1 shows the geometry schematically.

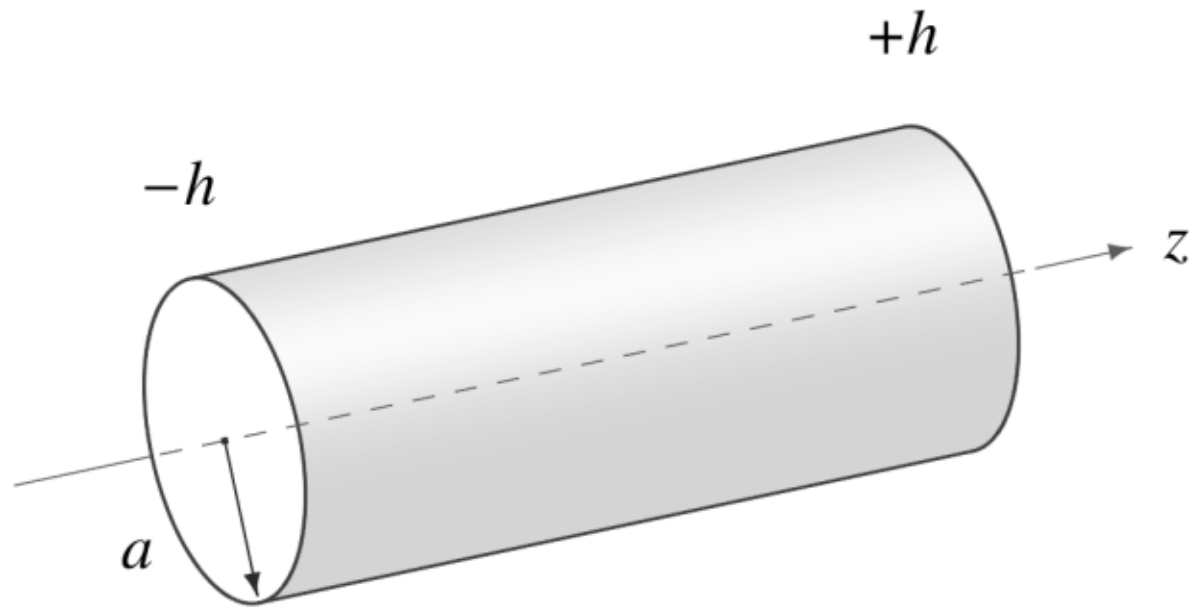

Figure 1. Geometry of the hollow metal cylinder.

The conductor is a circular cylindrical shell of negligible thickness, open at both ends, immersed in a homogeneous medium of permittivity $\varepsilon$ and held at potential $V$ relative to infinity. In the mid-1990s, the problem received a detailed analytical and numerical treatment. That treatment remained incomplete because the prevailing approach was to reformulate canonical problems of this kind for numerical solution. We extend the analysis and derive expressions for theoretical and practical calculations of the capacitance. The theoretical formulas involve the Meijer $G$-function. For practical calculations, we seek a simple formula that represents the capacitance over a wide range of aspect ratios $s = h/a$ while introducing an error of less than a few percent for every value of $s > 0$. Such a formula is useful in designing equipment and systems that contain hollow cylinders.

The paper comprises six sections. Section 2 briefly restates the problem, which the 1995 paper formulates in more detail. Section 3 presents two series expansions for the entries of the system matrix, in ascending and descending powers of the aspect ratio, respectively. Section 4 computes the capacitance, and Section 5 presents practical approximations for it. Section 6 draws conclusions and outlines future directions, particularly for the deliberate and effective use of artificial intelligence, which supported the developments in this paper through successive instructions and checks.

## 2. Problem Formulation

Consider a perfectly conducting cylindrical shell of negligible thickness, open at both ends and immersed in a homogeneous medium of electric permittivity $\varepsilon$. Choose coordinates with the $z$-axis along the cylinder axis and the origin at its center. The lateral surface is defined by

$$r = a\,, \qquad -h \leq z \leq h\,,$$

where $a > 0$ is the radius and $2h > 0$ is the total length of the cylinder. The conductor is held at a nonzero potential $V$ relative to infinity. By symmetry, the surface charge density $\sigma(z)$ is independent of the azimuthal angle and even in the axial coordinate. We extend it by zero for $|z| > h$, outside the lateral surface. The potential on the lateral surface is

$$\varphi(a, z) = \frac{a}{2\varepsilon}\int_{-h}^{h} \sigma(z_0) g(z - z_0)\, dz_0\,, \tag{1}$$

where the axisymmetric kernel is given by [2]

$$g(z) = \frac{1}{2\pi}\int_0^{2\pi} \frac{d\theta}{\sqrt{4a^2 \sin^2(\theta/2) + z^2}} = \frac{2}{\pi}\frac{K[4a^2/(z^2 + 4a^2)]}{\sqrt{z^2 + 4a^2}}\,, \tag{2}$$

Here $K(m)$ denotes the complete elliptic integral with parameter $m$. Imposing the equipotential boundary condition

$$\varphi(a, z) = V\,,$$

yields the Fredholm integral equation of the first kind

$$\frac{a}{2\varepsilon}\int_{-h}^{h} \sigma(z_0) g(z - z_0)\, dz_0 = V \quad \text{for} \quad |z| \le h\,. \tag{3}$$

The kernel has an integrable logarithmic singularity, with

$$g(z) = \frac{1}{\pi a}\ln\frac{8a}{|z|} + O\left[\frac{z^2}{a^2}\left(1 + \left|\ln\frac{|z|}{a}\right|\right)\right] \quad \text{as} \quad \frac{z}{a} \to 0\,.$$

Direct discretization is delicate because the equation is of the first kind and its kernel is singular. Reformulating the problem in terms of dual integral equations

[1] yields a Chebyshev expansion of the charge density,

$$\sigma(z) = \frac{2\varepsilon V}{\sqrt{h^2 - z^2}} \sum_{p=1}^{\infty} (-1)^{p-1} C_p T_{2p-2}(z/h) \quad \text{for} \quad |z| < h\,, \tag{4}$$

which factors out the correct edge behavior. The divergence at the rims is integrable and arises in the idealized model of negligible thickness. It indicates where charge accumulates and allows the potential and field to be reconstructed. In particular, the behavior near the rims matters when evaluating the local field. Predicting the peak field for a physical cylinder also requires the wall thickness and rim radius of curvature, which the model does not include.
The expansion coefficients satisfy the system

$$\sum_{p=1}^{\infty} D_{p,n} C_p = \delta_{n,1} \quad (n \in N^+)\,, \tag{5}$$

where $\delta_{n,1}$ is the Kronecker delta and

$$D_{p,n}(s) = D_{n,p}(s) = \frac{2}{s} \int_0^{\infty} I_0(u/s)\, K_0(u/s)\, J_{2n-2}(u)\, J_{2p-2}(u)\, du\,. \tag{6}$$

The coefficients $D_{p,n}$ and $C_p$ are dimensionless. For fixed $p, n$ and $s > 0$, the integral converges absolutely because the integrand is $O\big[u^{2(p+n-2)}(1 + |\ln u|)\big]$ at zero and $O(u^{-2})$ at infinity. Consequently, $D_{p,n}(s)$ is continuous on $(0, \infty)$. The physical capacitance $C = Q/V$ follows from the total charge,

$$C = \frac{2\pi a}{V} \int_{-h}^{h} \sigma(z)\, dz = 4\pi^2 \varepsilon a C_1\,. \tag{7}$$

The next section evaluates $D_{p,n}(s)$ and hence the matrix entries needed to determine the first expansion coefficient

$$C_1^{(N)}(s) = \frac{\begin{vmatrix} D_{2,2} & D_{2,3} & \cdots & D_{2,N} \\ D_{3,2} & D_{3,3} & \cdots & D_{3,N} \\ \cdots & \cdots & \cdots & \cdots \\ D_{N,2} & D_{N,3} & \cdots & D_{N,N} \end{vmatrix}}{\begin{vmatrix} D_{1,1} & D_{1,2} & \cdots & D_{1,N} \\ D_{2,1} & D_{2,2} & \cdots & D_{2,N} \\ \cdots & \cdots & \cdots & \cdots \\ D_{N,1} & D_{N,2} & \cdots & D_{N,N} \end{vmatrix}} \quad \to \quad C_1(s) = \lim_{N\to\infty} C_1^{(N)}(s)$$

and, from it, the capacitance

$$C(s) = 4\pi^2 \varepsilon a C_1(s) \,.$$

## 3. Entries of the Coefficient Matrix

To evaluate the integral (6) defining the matrix entries in the system, we first establish the two asymptotic behaviors

$$\lim_{s\to 0^+} D_{p,n}(s) = \begin{cases} +\infty \,, & p = n = 1 \,, \\ 1/[4(n-1)] \,, & p = n \geq 2 \,, \\ 0 \,, & p \neq n \,, \end{cases} \quad \lim_{s\to\infty} D_{p,n}(s) = 0 \,.$$

In particular, $D_{1,1}(s) \to +\infty$, with a logarithmic divergence as $s \to 0^+$, as shown below. The limit $D_{p,n}(s) \to 0$ as $s \to \infty$ does not imply that the capacitance tends to zero, since the first coefficient of the solution depends on the inverse of the full system matrix. For $s \to 0^+$, the boundedness under the condition $p + n > 2$ of $xI_0(x)K_0(x)$ justifies passing the limit under the integral sign. The case $(1,1)$ requires the logarithmic expansion given below.

■ We first examine the limit as $s \to 0^+$ for all entries except the case $p = n = 1$.

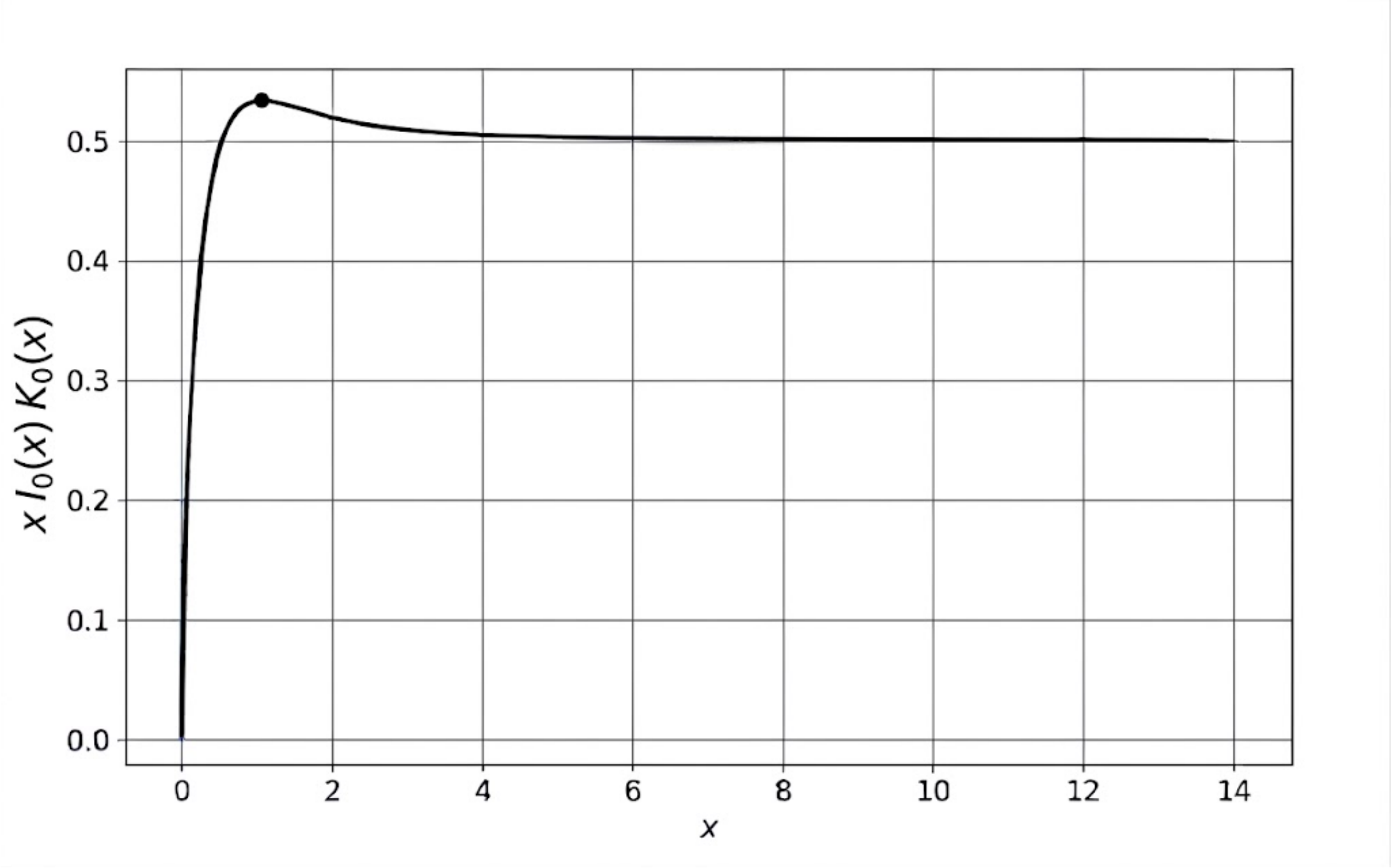

Figure 2. The function $y = xI_0(x)K_0(x)$.

The relevant asymptotic behaviors [2] are

$$\lim_{x\to 0^+} [xI_0(x)K_0(x)] = 0\,, \quad \lim_{x\to +\infty} [xI_0(x)K_0(x)] = \frac{1}{2}$$

and the function has a single local maximum,

$$\max[xI_0(x)K_0(x)] \cong 0.533 \quad \text{for} \quad x \cong 1.075\,.$$

It follows that

$$\lim_{s\to 0^+} D_{p,n}(s) = \int_0^\infty \frac{J_{2n-2}(u)\, J_{2p-2}(u)}{u}\, du\,,$$

which, by the Weber–Schafheitlin integrals [2], becomes

$$\lim_{s\to 0^+} D_{p,n}(s) = \frac{1}{2(p+n-2)\Gamma(1-|n-p|)\Gamma(1+|n-p|)} .$$

Furthermore, since

$$\frac{1}{\Gamma(1-|n-p|)} = \begin{cases} 1 , & \text{for } n = p , \\ 0 , & \text{for } n \neq p , \end{cases}$$

we obtain the first part of the desired result,

$$\lim_{s\to 0^+} D_{p,n}(s) = \begin{cases} 1/[4(n-1)] , & p = n \geq 2 , \\ 0 , & p \neq n . \end{cases}$$

In the special case $p = n = 1$, the integrand diverges at zero, and hence

$$\lim_{s\to 0^+} D_{1,1}(s) = \int_0^\infty \frac{J_0(u)\, J_0(u)}{u} du = \infty .$$

More precisely, as discussed below, one can show that

$$D_{1,1}(s) = \ln\frac{16}{s} - \frac{s^2}{16}\left(\ln\frac{16}{s} - 2\right) + O\left(s^4 \ln\frac{16}{s}\right) \quad \text{as} \quad s \to 0^+ ,$$

which gives a logarithmic divergence.

■ We next consider the second limit. Since

$$x I_0(x) K_0(x) \cong x\left(-\ln\frac{x}{2} - \gamma\right) \quad \text{as} \quad x \to 0^+$$

and the integrand can be bounded by an integrable function for $s \geq 1$, dominated convergence gives

$$\lim_{s \to \infty} D_{p,n}(s) = 0\,.$$

We now derive a closed-form representation of a general matrix entry in terms of a Mellin–Barnes integral,

$$D_{p,n}(s) = \frac{1}{2\pi} G_{5,5}^{2,4}\left( s^2 \middle| \begin{matrix} \frac{1}{2}, & \frac{1}{2}, & \frac{1}{2}, & 0, & \frac{1}{2} \\ 0, & r, & q, & -q, & -r \end{matrix} \right), \tag{8}$$

which uses the Meijer $G$-function, a general-purpose special function introduced by Cornelis Simon Meijer in 1936. For brevity, we set

$$r = p + n - 2\,,\ \ q = |p - n|\,.$$

Meijer originally represented the function by a series. Its modern definition uses a contour integral in the complex plane formulated in 1953 by the Hungarian mathematician Arthur Erdélyi.

To derive representation (8), consider the transform named after Robert Hjalmar Mellin for a general matrix entry,

$$\widehat{D}_{p,n}(\sigma) = \int_0^\infty D_{p,n}(s)\, s^{\sigma-1}\, ds \quad \text{with} \quad \sigma \in \mathbb{C} \quad \wedge \quad 0 < \mathrm{Re}(\sigma) < 1\,.$$

Substituting the integral definition (6) and interchanging the order of integration gives

$$\widehat{D}_{p,n}(\sigma) = 2\int_0^\infty J_{2n-2}(u)\, J_{2p-2}(u)\left[\int_0^\infty I_0(u/s)\, K_0(u/s)\, s^{\sigma-2}\, ds\right] du\,.$$

Since the integral in square brackets can be written as

$$\int_0^\infty I_0\left(\frac{u}{s}\right) K_0\left(\frac{u}{s}\right) s^{\sigma-2}\, ds = u^{\sigma-1}\int_0^\infty I_0(\theta)\, K_0(\theta)\theta^{-\sigma}\, d\theta\,,$$

the transform factors as

$$\widehat{D}_{p,n}(\sigma) = 2\int_0^\infty u^{\sigma-1} J_{2n-2}(u)\, J_{2p-2}(u)\, du \cdot \int_0^\infty \frac{I_0(\theta)\, K_0(\theta)}{\theta^\sigma} d\theta\,.$$

Evaluating the two integrals [2] in their common strip of convergence in the complex plane,

$$0 < \mathrm{Re}(\sigma) < 1\,,$$

and denoting the Euler gamma function by $\Gamma(z)$, we obtain

$$\widehat{D}_{p,n}(\sigma) = \frac{0.5\,\Gamma^2\left(\frac{1-\sigma}{2}\right)\Gamma(\sigma)\Gamma\left(p+n-2+\frac{\sigma}{2}\right)\Gamma(1-\sigma)}{\Gamma^2\left(\frac{1+\sigma}{2}\right)\Gamma\left(1-|p-n|-\frac{\sigma}{2}\right)\Gamma\left(1+|p-n|-\frac{\sigma}{2}\right)\Gamma\left(p+n-1-\frac{\sigma}{2}\right)}\,.$$

Mellin inversion,

$$D_{p,n}(s) = \frac{1}{2\pi i}\int_{c-i\infty}^{c+i\infty} \widehat{D}_{p,n}(\sigma)\, s^{-\sigma}\, d\sigma \quad \text{with} \quad 0 < c < 1\,,$$

then gives representation (8). For example, setting $n = p = 1$ yields

$$D_{1,1}(s)=\frac{1}{2\pi}G_{5,5}^{2,4}\left(s^2\left|\begin{array}{ccccc}\frac{1}{2}, & \frac{1}{2}, & \frac{1}{2}, & 0, & \frac{1}{2}\\ 0, & 0, & 0, & 0, & 0\end{array}\right.\right)$$

This representation can be evaluated and differentiated at any $s$. The resulting Taylor expansions, with decimal coefficients rounded to six places, have the form

$$D_{1,1}(s)=2.730939-1.024113(s-1)+0.533884(s-1)^2$$
$$-0.334950(s-1)^3+O[(s-1)^4]\,.$$

Figure 3 shows the logarithmic growth of $D_{1,1}(s)$ as $s\to 0^+$. The function is strictly decreasing and tends to zero as $s$ increases.

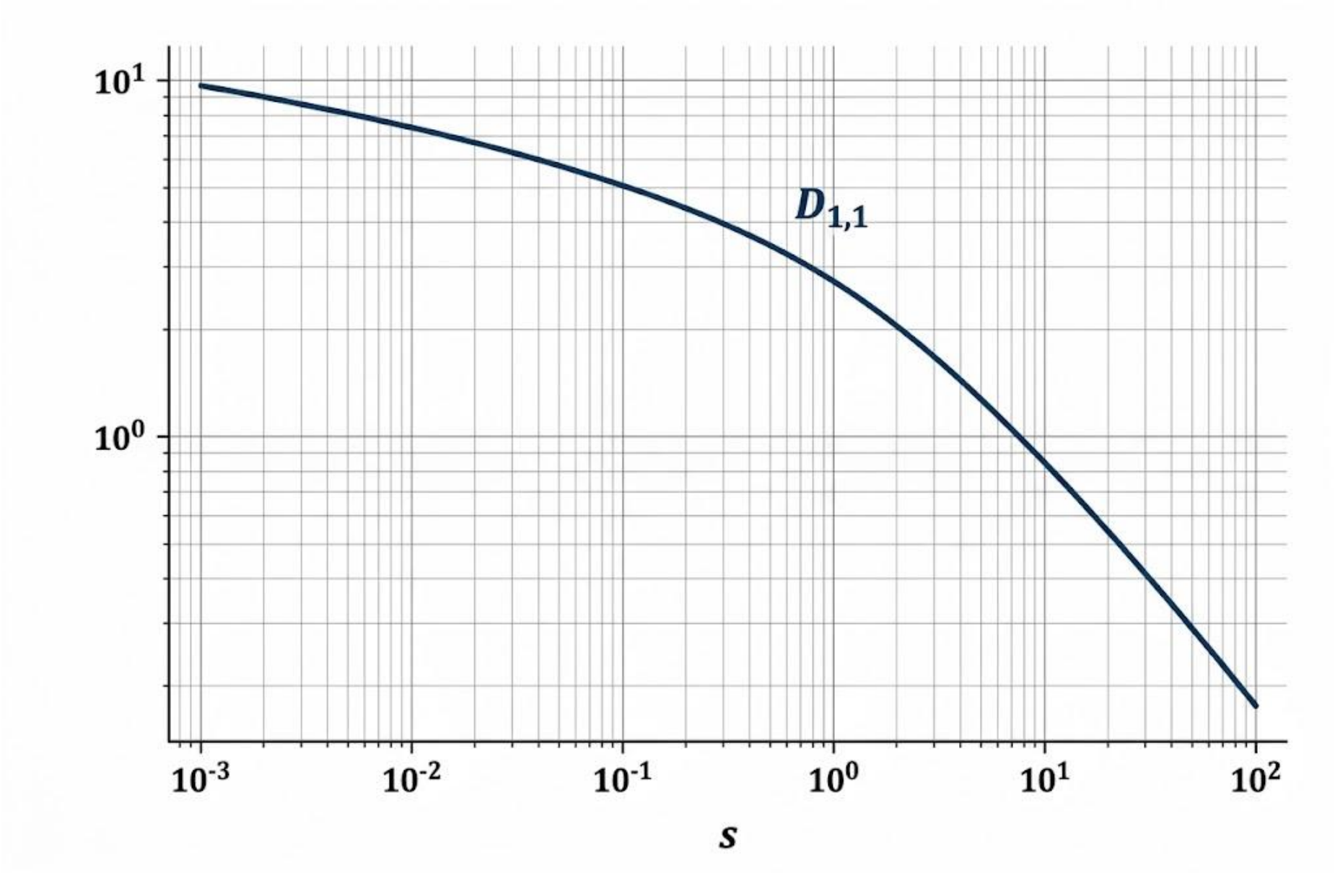

Figure 3. Behavior of the matrix entry $D_{1,1}(s)$.

This matrix entry is positive, as follows from its definition,

$$D_{1,1}(s) = \frac{2}{s}\int_0^\infty I_0(u/s)\, K_0(u/s)\, J_0^2(u)\, du > 0\,,$$

because the integrand is always positive. More generally, when $n = p$, the integrand is positive, so all main-diagonal entries are positive. Figure 4 shows two further entries to illustrate that an off-diagonal entry can change sign. The asymptotic estimates are

$$D_{1,2}(0) = 0\,,\;\; D_{1,2}(\infty) = 0\,,\;\; D_{2,2}(0) = \frac{1}{4}\,,\;\; D_{2,2}(\infty) = 0\,.$$

To interpret these curves, recall the Bessel identity [2],

$$\int_0^\infty J_{2n-2}(sx)\, J_{2p-2}(sx)\, \frac{ds}{s} = 0 \quad \text{for} \quad n \neq p\,,$$

which alone does not justify claiming that all entries $D_{p,n}(s)$off the main diagonal oscillate over $s > 0$.

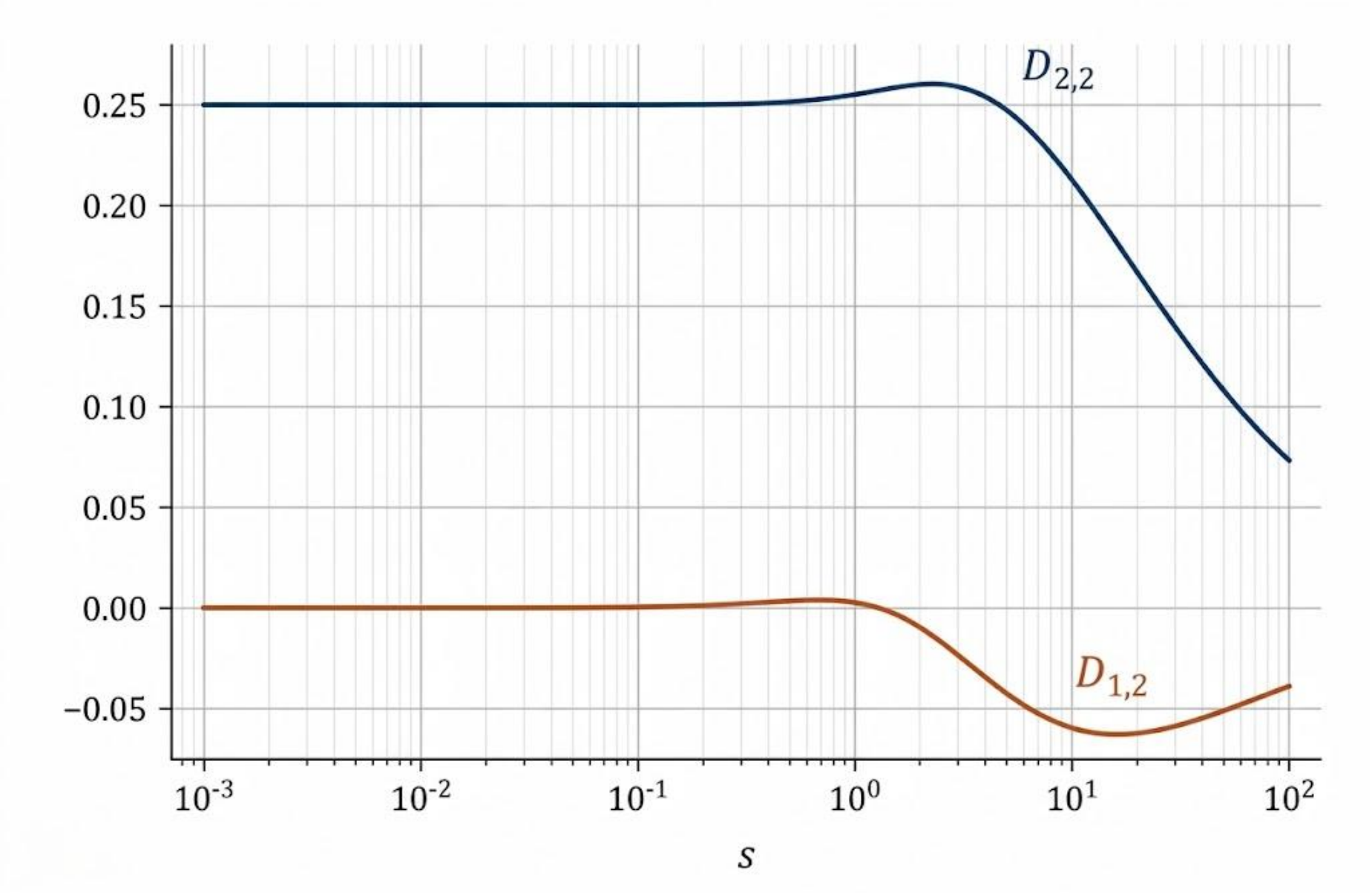

Figure 4. Behavior of the matrix entries $D_{1,2}(s)$ and $D_{2,2}(s)$.

For the entry shown in Figure 4, the following two asymptotic equivalents suffice:

$$D_{1,2}(s) = \begin{cases} \dfrac{s^2}{64}\ln(16/s) & \text{for } s \to 0^+ , \\ -\dfrac{\ln^2 s}{\pi s} & \text{for } \to \infty , \end{cases}$$

These equivalents are discussed below. The first is positive for sufficiently small $s$ and the second is negative for sufficiently large $s$. Continuity of $D_{1,2}(s)$ on $s > 0$ implies at least one zero between these regimes.

■ Finally, we give the asymptotic series for the entries of the coefficient matrix. The Meijer special function representation [3, 4] yields the series expansions below.

For $0 < s \le 1$, with $s = 0$ evaluated by taking the limit, we obtain

$$D_{p,n}(s) = \sum_{k=q}^{r-1} \alpha_k(r,q) s^{2k} + \ln\frac{16}{s} \sum_{k=r}^{\infty} \beta_k(r,q) s^{2k} + \sum_{k=r}^{\infty} \beta_k(r,q)\gamma_k(r,q) s^{2k} ,$$

where, for brevity, we set

$$\alpha_k(r,q) = \frac{1}{2}\frac{(1/2)_k^4\,(r-k-1)!}{(k-q)!\,(k+r)!\,(k+q)!},$$

$$\beta_k(r,q) = \frac{(-1)^{k+r}(1/2)_k^4}{(k-r)!\,(k-q)!\,(k+r)!\,(k+q)!},$$

$$\gamma_k(r,q) = 2H_k - 4H_{2k} + \frac{H_{k-r} + H_{k+r} + H_{k-q} + H_{k+q}}{2},$$

Here $H_k$ denotes the $k$ −th harmonic number, with $H_0 = 0$. The Pochhammer symbol is defined in terms of the Euler gamma function $\Gamma(x)$

$$(x)_n = \frac{\Gamma(x+n)}{\Gamma(x)}\,.$$

The first sum is empty if $r = q$.

For $s \geq 1$, the inversion identity

$$G_{p,q}^{m,n}\left(\frac{1}{z}\middle|\begin{matrix} a_1, & \cdots & a_p \\ b_1, & \cdots & b_q \end{matrix}\right) = G_{q,p}^{n,m}\left(z\middle|\begin{matrix} 1-b_1, & \cdots & 1-b_q \\ 1-a_1, & \cdots & 1-a_p \end{matrix}\right),$$

gives the following expression for a general matrix entry:

$$D_{p,n}(s) = \frac{1}{2\pi} G_{5,5}^{4,2}\left(\frac{1}{s^2}\middle|\begin{matrix} 1, & 1-r, & 1-q, & 1+q, & 1+r \\ \frac{1}{2}, & \frac{1}{2}, & \frac{1}{2}, & 1, & \frac{1}{2} \end{matrix}\right) \quad \text{for} \quad s \geq 1\,,$$

Equivalently, the series expansion is

$$D_{p,n}(s) = \frac{1}{\pi s}\sum_{k=0}^{\infty} \frac{(-1)^k}{s^{2k}} \left[\frac{(1/2)_k}{k!}\right]^4 \delta_k(r,q) Q_{r,q}(k,s)\,,$$

where, for brevity, we set

$$\delta_k(r,q) = (-1)^q \frac{(k+1/2)_r\,(k+1/2)_q}{(1/2-k)_r\,(1/2-k)_q},$$

$$\varepsilon_k(s) = \ln(16s) + 4(H_k - H_{2k})\,,$$

$$\zeta_k(r,q) = \frac{\pi^2}{12} + 2H_k^{(2)} - 4H_{2k}^{(2)} + \frac{S_r(k)+S_q(k)}{4},$$

$$S_m(k) = \sum_{j=0}^{m-1}\left[\frac{1}{(1/2-k+j)^2} - \frac{1}{(1/2+k+j)^2}\right],$$

$$Q_{r,q}(k,s) = \left[\varepsilon_k(r,q) - \frac{s_r(k)+s_q(k)}{2}\right]^2 + \zeta_k(r,q)\,,$$

$$s_m(k) = \sum_{j=0}^{m-1}\left[\frac{1}{1/2-k+j} + \frac{1}{1/2+k+j}\right].$$

For $m = r, q$, the sums $s_m(k)$ and $S_m(k)$ are empty if $m = 0$, and we define

$$H_k^{(2)} = \sum_{j=1}^{k} \frac{1}{j^2} \quad \text{with} \quad H_0^{(2)} = 0\,.$$

Both series converge absolutely even at $s = 1$, as can be checked numerically.

$$D_{p,n}(1^-) = D_{p,n}(1^+) = D_{p,n}(1)\,.$$

Although the summands differ, the two series sum to the same value. For fixed indices, the $O(k^{-2})$ decay of the terms gives absolute convergence, and the common integral representation gives equality at the joining point. Both series contain infinitely many terms, although their starting indices and summands can differ.

The table gives values at $p, n = 1, 2, 3$ rounded to ten decimal places. All entries are dimensionless. The negative sign of $D_{1,3}(1) = D_{3,1}(1)$ is correct because off-diagonal entries can be negative.

| $p/n$ | 1 | 2 | 3 |
|---|---|---|---|
| 1 | 2.7309391725 | 0.0026417630 | −0.0001980874 |
| 2 | 0.0026417630 | 0.2549796205 | 0.0007812119 |
| 3 | −0.0001980874 | 0.0007812119 | 0.1255872071 |

For $s = 1$, one could show that truncating the ascending and descending series gives the following inequality for their partial sums:

$$\left|S_M^{\uparrow} - S_{M-1}^{\downarrow}\right| \leq \frac{\mathrm{e}}{\pi^2}\left[\frac{1}{\left(M-\frac{3}{4}\right)^2} + \frac{1}{\left(M-\frac{1}{4}\right)^2}\right], \quad M \geq \max\{2, 2r+2, 2(r^2+q^2)\},$$

The inequality shows that the difference between the partial sums tends to zero as the number $M$ of terms increases. For the nine tabulated entries, the threshold $M \geq 32$ suffices for the bound to hold.

## 4. Calculation of the Capacitance

Once the entries of the coefficient matrix are known, the capacitance can be determined numerically. Truncating the matrix at successively higher orders gives the capacitance curve in dimensionless form, valid for any radius,

$$C_1(s) = \frac{C(s)}{4\pi^2 \varepsilon a} \quad \text{for every } s > 0\,. \tag{9}$$

Figure 5 shows the normalized capacitance $C_1(s)$ and the capacitance for a cylinder of base radius $a = 1\ cm$. At every plotted point, the maximum relative difference between the two truncations was below $1.1 \cdot 10^{-9}$.

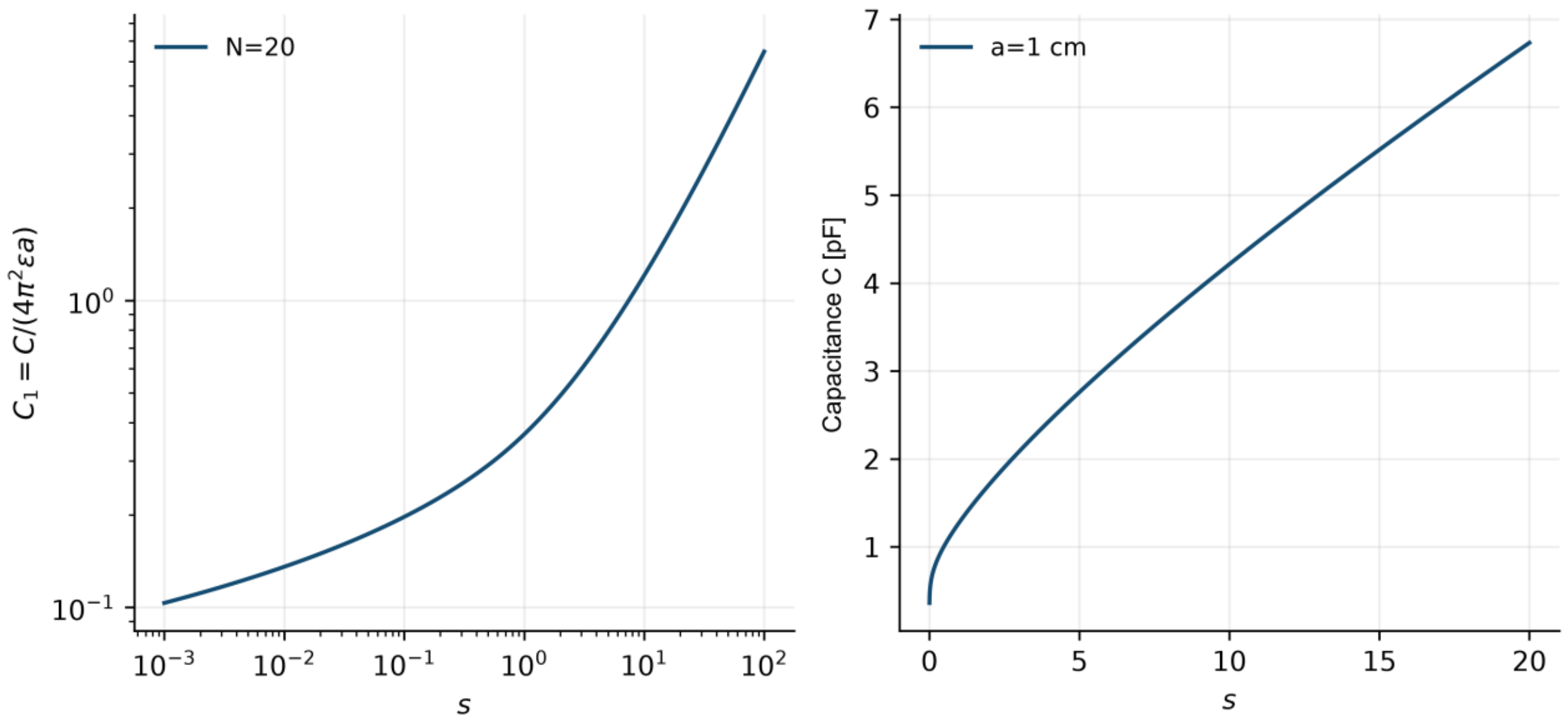

Figure 5. Dimensionless capacitance and an example of dimensional capacitance.

The truncation order is independent of the physical size of the tube. It helps determine the computational cost of obtaining a stable value. Agreement in capacitance alone does not guarantee the same accuracy for the field very close to the rims. The table illustrates convergence for $s = 100$, with $N$ denoting the matrix truncation order.

| $N$ | 1 | 4 | 12 | 20 | 30 |
|---|---|---|---|---|---|
| $C_1^{(N)}$ | 5.68576736 | 6.47142982 | 6.47192171 | 6.47192450 | 6.47192451 |

The relative difference between $N = 28$ and $N = 32$ is approximately $1.8 \cdot 10^{-12}$. Across the tabulated points, the maximum relative difference between $N = 20$ and $N = 32$ is $1.1 \cdot 10^{-9}$.

For a fixed radius and medium, the capacitance increases with $s$. A longer tube therefore requires more charge to reach the same potential. For small $s$, the capacitance decreases logarithmically as the aspect ratio decreases. At large aspect ratios, it grows more slowly than the length. The limiting formulas describe the ends of the curve, and the numerical solution of the system gives the intermediate regime.

■ The series in ascending powers for the matrix entries suffice to obtain the asymptotic expansions below by retaining successively more terms.

For small aspect ratios, $s \to 0^+$, we obtain

$$\frac{C(s)}{4\pi^2 \varepsilon a} = \sum_{k=0}^{\infty} a_k(L) s^{2k} \quad \text{with} \quad L = \ln \frac{16}{s} .$$

The first three coefficients of this expansion are

$$a_0(L) = \frac{1}{L} , \quad a_1(L) = \frac{L-2}{16L^2} , \quad a_2(L) = \frac{4L^3 - 85L^2 + 150L + 64}{4096L^3} ,$$

and hence

$$C_1 = a_0(L) + a_1(L)s^2 + a_2(L)s^4 + O(s^6) \quad \text{for} \quad s \to 0^+ .$$

For example, the first expansion coefficient follows from the leading approximation to the dominant matrix entry,

$$D_{1,1}(s) \cong \ln \frac{16}{s} \quad \to \quad C_1(s) \cong \frac{1}{\ln(16/s)} .$$

The remaining coefficients follow by repeating the calculation with more expansion coefficients. Higher-order terms can be obtained in the same way, although the formulas in the next section limit the practical need for them.

Equation (20) of the 1995 paper [1] contains a typographical error: replace $h/a$ with $a/h$. In the same paper, equation (21), which applies in the intermediate regime, should be read without the factor of two,

$$C_1(s) \cong \frac{\pi s}{\ln^2(16s) + \pi^2/12} .$$

This formula follows from truncation $N = 1$ and the leading term of $D_{1,1}$. For large aspect ratios, $s \to \infty$, we obtain

$$\frac{C(s)}{4\pi\varepsilon h} = \sum_{k=1}^{\infty} \frac{b_k}{\ln^k(2s)} ,$$

where the first three coefficients are

$$b_1 = 1 , \;\; b_2 = 1 - \ln 2 , \;\; b_3 = (1 - \ln 2)^2 + 1 - \frac{\pi^2}{12} .$$

For example,

$$\frac{C(s)}{4\pi\varepsilon h} = \frac{b_1}{\ln(2s)} + \frac{b_2}{\ln^2(2s)} + \frac{b_3}{\ln^3(2s)} + O[\ln^{-4}(2s)] .$$

Here too, higher-order terms can be obtained in the same way, although their practical value is limited given the global approximations in the next section.

Equation (22) of the 1995 paper [1] should be written as

$$\frac{C(s)}{8\pi\varepsilon h} = \frac{1}{\Omega} + \frac{4 - \pi^2/3}{\Omega^3} + \frac{16 - 10\zeta(3)}{\Omega^4} + \frac{144 - 8\pi^2 - 17\pi^4/45}{\Omega^5} + O(\Omega^{-6}) .$$

where $\Omega = 2[\ln(4s) - 1]$ and $\zeta(3)$ is Apéry's constant,

$$\zeta(3) = \sum_{n=1}^{\infty} \frac{1}{n^3} \cong 1.20205 \,.$$

The leading orders of this formula are consistent with Jackson's formula [5].

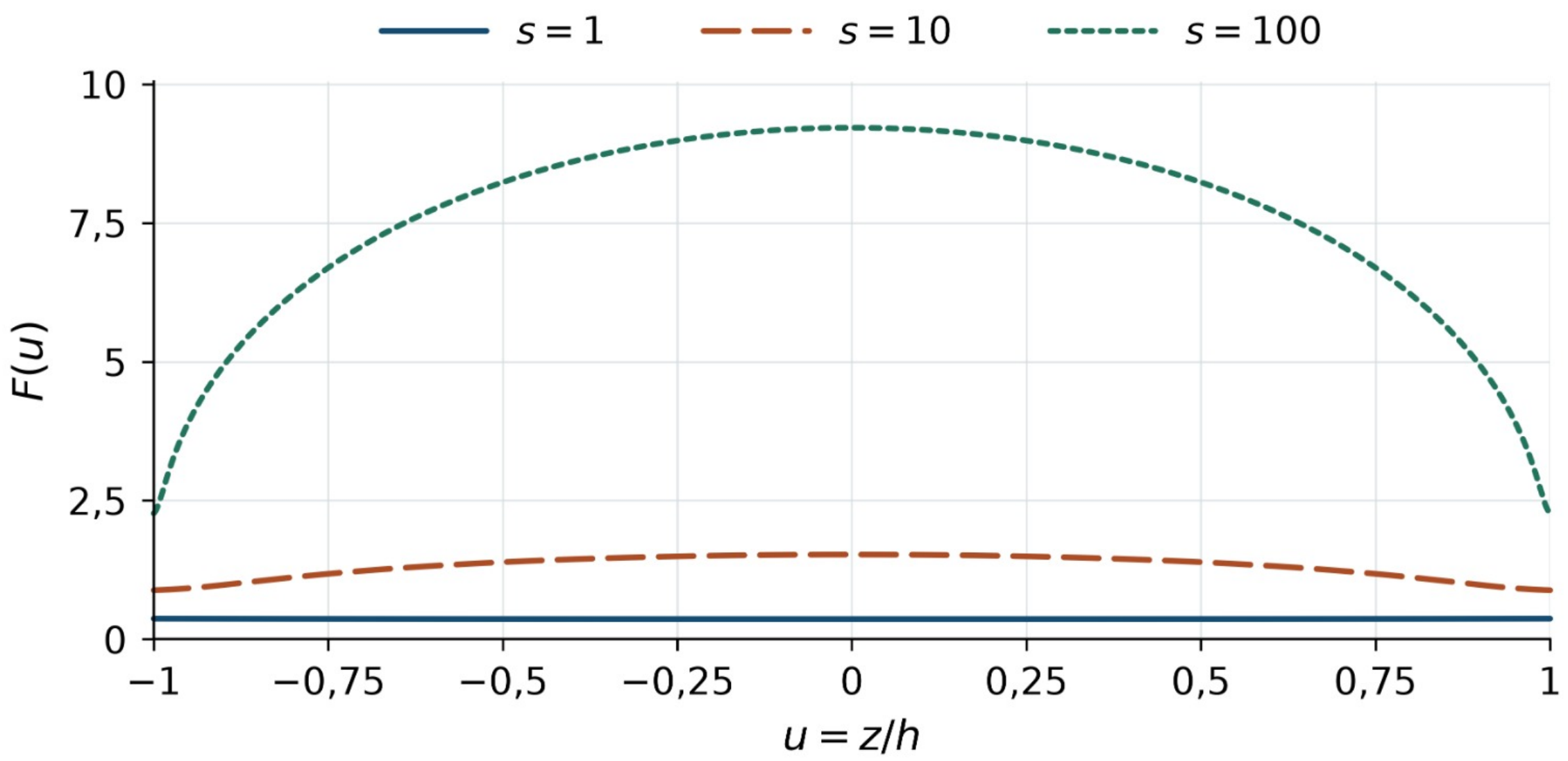

Figure 6. Variation of $F(u)$ along the metal cylinder.

Figure 6 illustrates the charge density along the cylinder for several aspect ratios. For clarity, it plots the dimensionless quantity below, from which the edge singularity has been factored out.

$$F(u) = \sum_{p=1}^{\infty} (-1)^{p-1} C_p T_{2p-2}(u) \quad \text{with} \quad u = \frac{z}{h},$$

This is the regular factor of the charge density (4) after removing the singular factors at the rims.

## 5. Global Approximation Formulas

To obtain an approximation valid for every $s > 0$, we first give three asymptotic expressions for the normalized capacitance,

$$C_1(s) = \frac{1}{\ln(16/s)}, \qquad \text{for a short cylinder,}$$

$$C_1(s) = \frac{s}{\pi \ln s}, \qquad \text{for a long cylinder,}$$

$$C_1(s) = \frac{s}{\pi[\ln(4s) - 1]}, \qquad \text{and for a uniformly charged cylinder.}$$

The first two formulas follow from the preceding analysis. The third is a practical estimate based on uniform charge and is also asymptotically equivalent in the long-cylinder limit. For example, the short-cylinder formula alone has an error of about 0.038% at $s = 0.1$, 0.625% at $s = 0.5$, and 1.50% at $s = 1$. The uniform-charge estimate has an error of about 1.84% at $s = 10$ and 1.47% at $s = 100$. These are numerical comparisons at a finite set of points. First specify the required tolerance, for example 1%, and then check whether the approximation meets it for the geometry of interest. If it does not, use the full system. In general, a valid asymptotic formula may approach its limit slowly.

The asymptotic expansions of the capacitance and the available numerical simulations yield a semianalytical formula that estimates the capacitance with a maximum error of 2%. Using the formula

$$\frac{C_{\text{app}}(s)}{4\pi^2 \varepsilon a} = \frac{1 + As}{\ln(1 + 16/s) + \pi A \ln(1 + Bs)},$$

one can readily check that it reproduces both asymptotic forms,

$$\frac{C(s)}{4\pi^2\varepsilon a} \cong \begin{cases} \dfrac{1}{\ln(16/s)}, & \text{for } s \to 0^+ , \\ \dfrac{s}{\pi \ln s} , & \text{for } s \to \infty . \end{cases}$$

Choosing

$$A = \frac{74}{125} = 0.592 , \quad B = \frac{1241}{1000} = 1.241 ,$$

gives a relative error of

$$e(s) = \frac{\left|C_{\text{app}}(s) - C(s)\right|}{C(s)} < 0.02 \quad \text{for every } s > 0 .$$

Over the interval $10^{-3} \leq s \leq 10^2$, comparison at 252 distinct points gave a maximum observed error of 1.891% on that sample. At this level of relative error, a plot of the approximate capacitance would be indistinguishable from the one already shown. The formula is useful when the cylinder capacitance must be estimated quickly and in more demanding simulations in which the metal cylinder is only one part of the system.

One can seek an approximation with a smaller error. The following formula, valid for $s > 0$,

$$\frac{C_{\text{app}}(s)}{4\pi^2\varepsilon a} = \frac{1+s}{\ln(1+16/s) + \pi \ln(1+7s/5)} \left[1 - \frac{2s}{5(1+5s)^2} + \frac{4s}{5(s+4)^2}\right]$$

which gives capacitance estimates with an error of

$$e(s) = \frac{\left|C_{\mathrm{app}}(s) - C(s)\right|}{C(s)} < 0.01 \quad \text{for every } s > 0 \, .$$

The expression, however, is more complicated and less convenient to use.

## 6. Conclusions and Outlook

We revisited the electrostatic capacitance of a hollow metal cylinder by extending the formulation proposed in 1995 [1] and connecting the analytical representations to a practical expression. Under the assumptions of the model, the approximation formulas give the capacitance through elementary operations with a relative error no greater than $(1 \div 2)\%$ for every aspect ratio $s > 0$. Deriving these expressions required comparing the limiting behaviors, analytical expansions, and numerical results. We examined and corrected successive approximations to balance simplicity and accuracy. Estimating the error is an essential part of the result because it establishes how the formula can be used and defines its reliability in applications.

Our approach was inspired by the work of Giorgio Parisi and Francesco Zamponi [6]. They describe the contribution of artificial intelligence to developing an analytical proof, alongside the verification of individual steps, correction of inconsistencies, and revision of the text. Their experience suggests a practical way to integrate AI tools into theoretical research while subjecting the outcome to the usual standards of scientific verification.

In this context, the analogy of AI as a "student to be taught" is useful. Teaching here means providing the problem, assumptions, conventions, and available results, making precise requests, and checking the responses in successive rounds [7]. The quality of the interaction also depends on the researcher's ability to spot errors, clarify intermediate steps, and steer the work toward alternatives that can be tested. In every case, the authors remain responsible for the conclusions.

The use of AI thus opens a new direction in scientific research by allowing researchers to explore further representations of classical problems and quickly compare candidate solutions. Possible extensions of this work include more accurate approximations and a treatment of cylinders of finite thickness or cylinders in the presence of other conductors. These extensions will require a new formulation of the model and an error assessment specific to each case. Documenting the analytical steps, computations, and contributions of AI will be essential for reproducibility.